# ANALYSIS OF IPv6 TRANSITION TECHNOLOGIES

Ali Albkerat and Biju Issac

School of Computing, Teesside University, Middlesbrough, UK

*ABSTRACT*

*Currently IPv6 is extremely popular with companies, organizations and Internet service providers (ISP) due to the limitations of IPv4. In order to prevent an abrupt change from IPv4 to IPv6, three mechanisms will be used to provide a smooth transition from IPv4 to IPv6 with minimum effect on the network. These mechanisms are Dual-Stack, Tunnel and Translation. This research will shed the light on IPv4 and IPv6 and assess the automatic and manual transition strategies of the IPv6 by comparing their performances in order to show how the transition strategy affects network behaviour. The experiment will be executed using OPNET Modeler that simulates a network containing a Wide Area Network (WAN), a Local Area Network (LAN), hosts and servers. The results will be presented in graphs and tables, with further explanation. The experiment will use different measurements such as throughput, latency (delay), queuing delay, and TCP delay.*

## 1. INTRODUCTION

The connection between computing nodes requires a protocol, number or a name, in order for each node to be recognized, and for the source and destination of each packet to be known. The Internet depends on protocol that is known as Internet Protocol version 4 (IPv4), which uses Classless Interdomain Routing (CIDR) and 32 bit: this protocol can cover 4.3 billion nodes around the world. Because the technology is developing, and many different services and devices use 3G and 4G, IPv4 is approaching its limit: there is not enough IPs available from internet service providers (ISP) to meet customer demand. Therefore, the new version of IP is critical in maintaining the pace of the Internet's development as shown in the picture below. IPv6, developed by IETF, is considered more efficient than IPv4 in relation to scalability, reliability, speed and security. Moreover, the size is larger than IPv4, as it uses 128 bit that will be able to encompass all of the nodes and any services that might require the IP, both now and in the future. Countries such as China, India and Japan have begun to use the next generation IP
[1].

IPv6 can cover 340 trillion, trillion, trillion nodes whereas IPv4 is only capable of 4.3 billion nodes. This will contribute to building the necessary infrastructure for future development. IPv6 will not require NAT as IPv4 does, as security will be built in. IPv4 used NAT as security, but its function is not primarily for security. The flow control provides high priority for specific traffic to avoid congestion, and the connection with IPv6 will be as end to end. In addition, the IPv6 header is simpler than IPv4. It contains fewer fields which helps data to be processed faster, which will in turn be reflected in a higher performance. One of the fields which will not be included with IPv6 is the CRC, because the packet is already checked at a lower layer, and therefore is not required to be checked for errors in an upper layer. Consequently the process time will be decreased. The transition from IPv4 to IPv6 requires a smooth method without any disconnection or fault within the network. This requires an efficient management and upgrading of the nodes, devices and operation systems in order to be able to understand the new IP generation.

## 2. UNDERSTANDING IPv4 AND IPv6

The communication evaluation identifies IPv4 as being limited in not only the addresses available for customers but also in the services that consumers need to access the applications. The new version (IPv6) is found to have solved these issues of IPv4 by extending the size of the network in order to accommodate more customers; it is also easier to reconfigure addresses. IPv6 also provides a higher performance, particularly during real time traffic, which requires quality of service (QoS), and the overall processing time is reduced. Moreover, the new version is able to provide what is required for future development of an infrastructure. Security is also an important consideration, as the internet is used by many different applications to transfer data; security is implemented with the IPv6. In addition, mobility is supported.

### 2.1. IPv4

IPv4 is considered the core of internet addressing, as it allows transmission of data using TCP/IP. In previous years, this protocol proved its stability and reliability in working in the internet environment in order to provide a connection for millions of nodes. Figure 1 shows the five classes of IPv4 address.

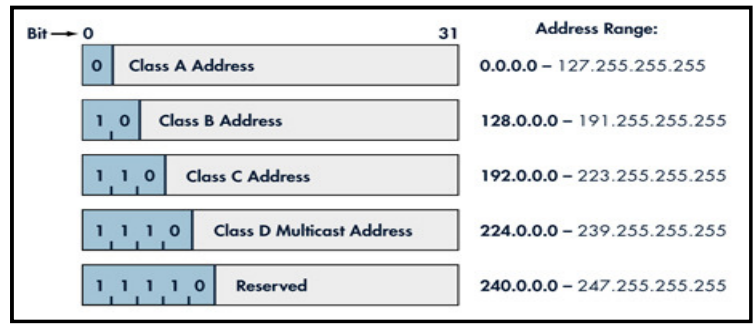

Figure 1. IPv4 addressing

IPv4 was launched in the 1980s. After a small period of time, this protocol started to be exhausted; this led to the use of class inter domain routing (CIDR); however, this did not provide a long-term solution due to the rapidly increased use of the Internet. Some sources expected it to be exhausted in 2010 or 2012, which was a primary reason to develop a new version that was able to accommodate more consumers.

IPv4 contains 32 bits. It can cover 4.3 billion addresses. The address is represented as 192.168.2.1. Each colon can be from 0 to 255. In general, IPv4 contains five classes. Each class provides different limits to the address numbers for networks and hosts; the figure 2 shows the types of addresses and their range.

## 2.2. IPv6

IPv6 is also known as IP next generation: it is considered evolutionary from IPv4, as it does not make a radical change to IPv4 and the basic concept remains the same, but some features have been added, which help to improve performance and provide a good service for customers. In IPv6, the NAT was eliminated, which is considered an advantage. Moreover, the configuration is easier with IPv6 as it can be done stateless (auto configuration). The IP address is a combination of the MAC address for the interface and the prefix from the router; in general the DHCP is not used, but it can be used with DNS. The IPv6 size is 128 bits, comprised of Hexadecimal digits which are able to provide $3.8 \times 10^{38}$ addresses, which are enough to give a unique address to each device for today and the future. Each four digits are separated by a colon which provides eight parts; the zeroes can be omitted to make the address smaller as shown in figure 2 [2].

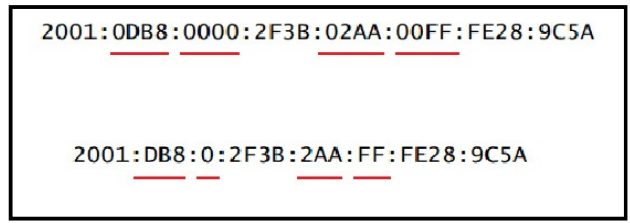

Figure 2. IPv6 addressing

IPv6 makes the global routing simpler than IPv4. There is less effect on resources and memory, which helps to improve performance and be more efficient. The security is provided end to end by encryption, which is integrated within IPV6.

The traffic in IPv4 can be unicast, multicast or broadcast. With IPV6 the broadcast is no longer available because of its high consumption of resources. However, a broadcast can be sent from within a multicast.

1. Unicast Addressing: the unicast is to send the packet for a unique address (for one destination).
2. Multicast Addressing: the multicast addressing is to send the packet for a group of addresses. The IPv6 use the ff00::/8 as a prefix for multicast. The type of addressing use two protocols to know which IPs in the same group for multicast there are Multicast Listening Discovery protocol (MLD) and MLDv2.
3. Anycast Addressing: when there are many similar destinations in different areas the unicast use to send the packets to the closet destination from the sender [3].

## 2.3. Header

The IPV6 header is quite different from IPv4. Although IPv6 has abolished some fields it is still bigger than IPv4; however, it is more efficient than IPv4. The picture below shows the two headers and the differences between them. For example, the CRC is no longer necessary as the packet is already checked from a lower layer so there is no reason to check it again, which makes overheads on processing and loses time. Figures 3 and 4 show the differences header for IPv4 and IPv6.

| Version | Traffic Class | Flow Label | |
|---|---|---|---|
| Payload Length | | Next Header (48-55) | Hop Limit |
| Source Address | | | |
| Destination Address | | | |

Figure 3. IPv6 header

| Ver | IHL | Type of Service | Total Length | |
|---|---|---|---|---|
| Identification | | | Flags | Fragment Offset |
| Time to Live | | Protocol | Header Checksum | |
| Source Address | | | | |
| Destination Address | | | | |
| Options + Padding | | | | |

Figure 4. IPv4 header

The IPv6 header is bigger than IPv4: IPv6 is 40 bytes while IPv4 is 20 bytes; this is how the IPv6 address is bigger than in IPv4. The version field detects which type of header it is: if it is 6 then the IP is IPv6, and if it is 4 that mean that it is IPv4. Traffic class determines the priority for traffic, which varies from 0 to 7: traffic class length is 8 bits and it is used to reduce traffic congestion as much as possible. The quality of service is provided by the flow label, which is 20 bits; when the traffic reaches the router this field provide a mechanism to process the traffic. Payload length, which is 16 bits, is used to detect the length of data and is able to transfer up to 64 Kbytes. The extension header will be used when the data exceeds 64 Kbytes, as it is 32 bits: the extension header therefore is capable of providing 4.3 million bytes. The type of extension header used is detected by the Next Header field. Hop limit is similar to TTL in IPv4: it is decreased after each hop until it reaches zero, and then it is discarded. The source and destination are each 16 bytes, which enables them to provide a long address [4].

Overall the IPv6 provides solutions for weakness in IPv4, such as address exhaustion: IPv6 provides addresses with 128 bit, there are no private addresses, and the transmission of data is end to end. IPv4 depends on manual or dynamic host configuration for addressing, whereas IPv6 uses auto configuration: the configuration is done automatically without the need to send a query and wait for a response from the DHCP server. The security with IPv4 is optional, so data transferred over the Internet could be hacked; IPv6, however, has IPsec in-built so that data is encrypted. IPv4 is limited in real time traffic despite using type of service (TOS) whereas IPv6 supports real time traffic by using Traffic Class and Flow Label. The routing table in IPv4 is large in comparison to IPv6. IPv6 uses the same home IP for excellent mobility, even when outside the home and it can use neighbour discovery and auto configuration when moving from one link to another [5].

## 3. TRANSITION STRATEGIES

Transition strategies are methods that provide a means of connection between IPv4 and IPv6, as these two protocols cannot understand each other. Therefore, in order to transfer data, a special method is needed. The three strategies are:

- Dual-Stack: This method is used to understand simultaneously IPv4 and IPv6: regardless of which protocol is used, when the traffic is received the node is able to respond.
- Tunnel: This strategy is employed when there are two networks that are using the same IP version but are separated by another network that has a different IP. The tunnel method establishes a virtual link through the networks by providing a connection in the middle of them.
- Translation: This method is similar to NAT, as it changes the IP packet from IPv4 to IPv6 and vice versa, depending on the source and the destination [6].

### 3.1. Dual-Stack

The Dual Stack technique uses IPv4 and IPv6 within the same stack in parallel. The choice of protocol is decided by the administrator policies, along with what kind of service is required and which type of network is used. This technology does not make any change to the packet header and at the same time does not make encapsulation between IPv4 and IPv6. This technology is known as native dual stack or Dual IP layer [7].

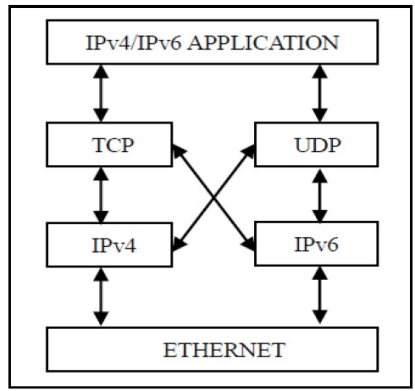

Figure 5 Dual-Stack

According to [8], the Internet contains nodes and these nodes are able to support both protocols in parallel within the same infrastructure. Therefore, the node can provide the transmission of data for IPv4 and IPv6. This technique is not suitable for large networks like the Internet because it is difficult and expensive to cover all the nodes in such huge networks. On the other hand, it is suitable for small networks, which need less management and are easy to control. The dual stack is considered to be the basis for inventing the two other techniques for transition between IPv4 and IPv6

## 3.2. Tunneling

Tunneling could be either manual or automatic. The connection for the manual is a point to point mode which is assigned the source and the destination address of the tunnel by the operator while the automatic connection is a point to multipoint where the source address is assigned by the operator and the destination address is found automatically. The tunnel idea works as a bridge to transfer packets between two similar networks over incompatible network [7]. In other words, the IPv6 will be as a part of IPv4, and the IPv6 data will flow by using IPv4 infrastructure, which will send it to the destination (IPv6) for processing; the tunnel is a virtual link between the two points to transfer data [9].

### 3.2.1. Manual Tunnel

The manual tunnel provides a connection between the IPv6 networks over the IPv4 network as a static point-to-point tunnel. The IPv4 and IPv6 are manually assigned as the source and destination. This strategy provides a secure connection between two ends [10].

### 3.2.2. Automatic Tunnel

There are different types of automatic tunnels as follows.

### 3.2.2.1. Tunnel Broker

The dual stack is important for a tunnel broker, so that a tunnel for the hosts in the IPv4 network only can be built. The web server is required to build the tunnel because the user should be connected to a web server and apply certain authentication details (such as the IP address, operating system and IPv6 support software) and the replay will be a short script; now the IPv4 to IPv6 tunnel is ready to use. The tunnel broker is considered to be an automatic configuration service and it will configure the end point for the network side, the DNS server and the end user [11]. The tunnel broker contains different parts: the first is the tunnel broker (TB), which sends instructions between the server and DNS. Additionally, the TB works as a tunnel monitor, and if the tunnel is down it can use other tunnels which are already in existence in the tunnel group. The second is the tunnel server (TS), which should have at least three IPv6, an IPv4 unicast, and an anycast: these are used for routing, accessibility, and endpoint for the user respectively. Third is the tunnel server group (TSG), which uses the IPv4 anycast to divide the tunnel servers into tunnel server groups, all of which have the same IPv4 anycast address. This makes tunnel work more efficient as the user's request will be sent to the nearest tunnel, and if there is any issue with the connection another tunnel will take over and generate the connection. The fourth part is the DNS system (DNS), where each user has a domain name and the mappings are done by the DNS system. This requires the user to register to access the tunnel, and then the user will obtain an IPv6 address; at the end, the communication is carried out by a website such as http://gogo6.com/ by using HTTP protocol [12], figure 6 shows the tunnel broker mechanism.

International Journal of Computer Networks & Communications (IJCNC) Vol.6, No.5, September 2014

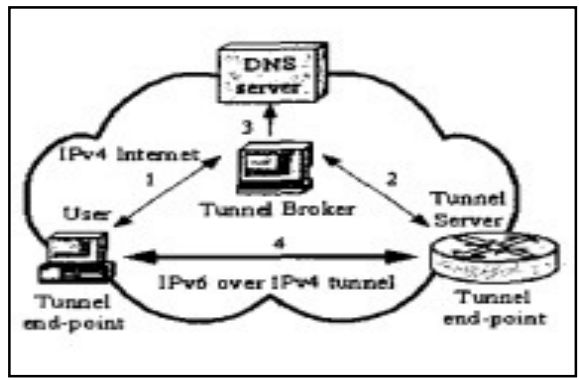

Figure 6. Tunnel Broker mechanism [11]

### 3.2.2.2. 6to4

6to4 is a technique that is able to connect IPv6 domains that are separated by an IPv4 network. The IPv4 network acts as a link between the IPv6 networks. 6to4 is an automatic tunnel. It uses the IPv4 infrastructure to transfer the IPv6 packet. Therefore, the IPv4 address is part of the IPv6 address during the transferring of the packets until they reach the other side of the tunnel [11]. The IPv6 networks are connected together using the 6to4 router with the prefix 2002: IPv4 address::/48. The IPv4 address (32 bit) is the 6to4 router address. The IPv6 destination will extract the encapsulation address. In addition to connecting the IPv6 network with the IPv6 Internet through the IPv4 network, the prefix is the same and the 6to4 router will encapsulate the IPv4 destination for the 6to4 relay router, as shown in figure 7, there are two IPv6 hosts isolated by the IPv4 network; the tunneling used by IPv6 to deliver data through IPv4 [9].

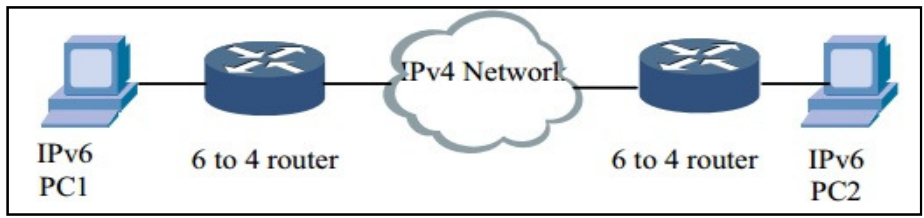

Figure 7. 6to4 mechanism [8].

### 3.2.2.3. 6over4

The 6over4 is an automatic technique for providing an approach to IPv6 nodes that exist within a pool of IPv4 networks. These IPv6 nodes are not directly connected to each other; so this technique will create a virtual link to provide a way for the IPv6 nodes to communicate [11]. The virtual link was created by IPv4 Multicast; it is represented by Ethernet with IPv6 and Multicast with IPv4. Therefore, the IPv4 infrastructure should be fully supported by IPv4 to provide the virtual link to all the IPv6 nodes. There are two important protocols to use with this technique, SLAAC and ND, the latter of which causes a security issue because the ND message might be attacked [8].

### 3.2.2.4. ISATAP (Intra-Site Automatic Tunnel Addressing Protocol)

ISATAP is another mechanism for enabling communication between IPv6 and IPv4 using the tunneling technique. It is used to link the local IPv6 address with the prefix fe80::5efe/96, which

25

is followed by the IPv4 32 bit. ISATAP can build more than one gateway, which is used as a tunnel for IPv6 to access ISATAP hosts [8].

ISTAP is an automatic tunnel and it is point to point connection. The addressing is dependent on embedding strategy; the IPv6 address will be within an IPv4 address. The ISATAP tunnel is able to provide a connection between IPv6 and IPv4 routers: at the beginning of the connection the host within ISTAP will get an address called a local ISATAP address and will detect the next hop of the ISATAP router. The packets will then be sent by the tunnel after embedding the IPv6 address into an IPv4 address. At the destination the IPv4 header will be removed and the packet is sent to the IPv6 server; there the server sends the packets to the ISATAP network and finally the ISATAP router prepares the IPv6 packets into IPv4 and sends them to the ISATAP host, which will then remove the IPv4 header and extract the IPv6 packets [13]. Figure 8 shows the ISATAP mechanism.

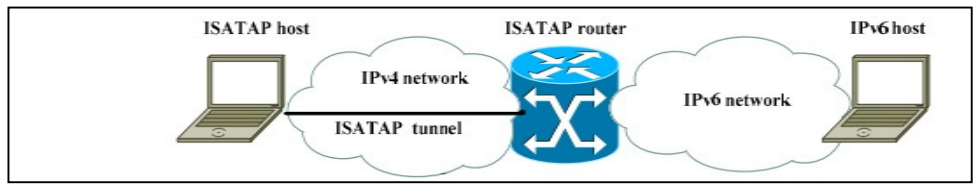

Figure 8. ISATAP mechanism (Xiaodong,2009)

## 3.3 Translation

The translation technology changes the header and the payload of the IP from version 4 to version 6 and vice versa. There are two ways for translation: stateless and stateful. The stateless translation, there is no reference for the pervious packet during the conversion while the stateful translation is related with the previous packets [7].

### 3.3.1 SIIT (Stateless IP/ICMP Translation)

The translation is executed with the header between IPv4 and IPv6. During the translation the information might be lost and NAT (network address translation) is required; therefore this technique is not recommended [14].The SIIT technique requires each IPv6 host to have an assigned IPv4 address. There are two types of addressing: one is known as IPv4-translated address for the IPv6 host, where the IPv6 address is generated by adding the prefix 0:ffff:0:0:0/96 before the IPv4 address; the second type is known as IPv4-mapped address for the IPv4 host, and the IPv6 is generated by adding ::ffff:0:0/96 before the IPv4 address. The translation operation is as follows: the IPv4 packet is translated to IPv6, the source will take the prefix ::ffff:0:0/96 and the destination will take the prefix 0:ffff:0:0:0/96 and remove it from the original. The DNS is vital in knowing the addresses; the local DNS server helps the IPv6 host to learn the IPv4 mapped address in order to get 'AAAA' record from 'A' record by using DNS64.

Moreover, the IPv4 record is registered within IPv6 hosts to answer the heterogeneous query and there is no security issue added for the network by the SIIT technique; also the DHCPv6 and the SLAA can be used to assign the IPv6 addresses for the host. This type is a stateless translation [8].

### 3.3.2 NAT-PT (Network Address Translation--Protocol Translation)

The communication between native IPv6 and native IPv4 could be obtained by using NAT-PT. This mechanism has a pool of global IPv4 and IPv6 prefix with length of 96 bits. The translation will be created by assigning the IPv6 with the IPv4 address pool through NAT-PT gateway [6].

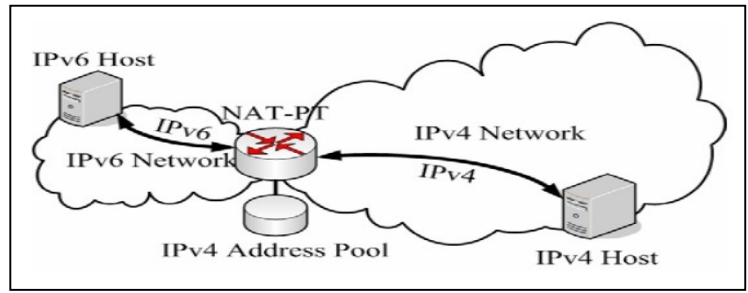

Figure 9. NAT PT [15].

This mechanism does not require extra Applications or depend on another mechanism, such as dual stack, but it requires interoperability with the core network for easy and fast management [11].

The prefix ::/96 will be used to generate a new address. To translate from IPv6 to IPv4, the IPv4 source will be created from the source of IPv6 and the port is found by looking it up in the NAT binding table; the destination IPv4 is created by removing the prefix. To translate from IPv4 to IPv6 the prefix will be added to the IPv4 source address to create an IPv6 source, and the destination address is generated by using the destination IPv4 address and the port looked up in the NAT binding table. To avoid the problems that are generated by building the binding map, the heterogeneous addressing will use the DNS ALG on the translator. This will assist in converting the A to AAAA query in two-way to generate a stateful binding between IPv6 and IPv4 addressing IPv4 by using the pool of addressing [8].

### 3.3.3 BIS (Bump in the stack) and BIA (Bump in the API)

Both BIA and BIS are stateful translations. These two mechanisms are used to solve a problem when an application in IPv4 wants to communicate with a remote IPv6 host through an IPv6 network; this strategy depends on tricking the application using IPv4 to assume that the remote host is IPv4 as well. This technique is built by software and inserted inside the host. The security is lax enough for a DOS attack on the DNS query: by exhausting the pool of IPv4 addresses, the binding table will be full [8]

### 3.3.3.1 Bump in the stack (BIS)

BIS uses the translation as per packet: the translation executes the operation by generating the source address from the host and the destination from the binding table with an IPv4 destination address. When the packet reaches the host, the translator translates the packet to IPv4 and the source address is taken from the binding table with the IPv6 source address and the destination from the host IPv4 address, as shown in figure 10 [8].

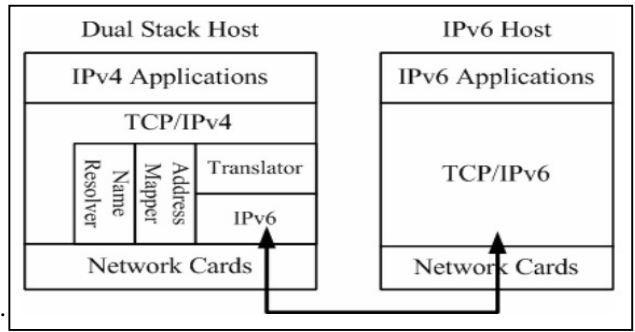

Figure 10. BIS [8]

The translation between IPv4 and IPv6 is done by injection BIS in the dual stack for the host and the IPv4 is detected to the IPv6 [15].

**3.3.3.2 Bump in the API (BIA)**

BIA is similar to BIS; with BIA the translator translates between IPv4 APIs and IPv6 APIs. The name resolver and address mapper are the same as in BIS, and the function mapper is responsible for the translation. The translation will be done without the IP header so that the security will not break down between end and end.

## 4. IPv6 DEPLOYMENT

To start using IPv6 over a network the equipment such as router, switches firewalls and servers should be supported by IPv6, and then the transition may begin. Generally, the migration will take time; however many countries today began to use IPv6 side by side with IPv4.

In Asia companies have begun to use IPv6 as IPv4 is no longer available. Governments in Japan, Korea and China prompt their countries to move to IPv6. The migration in Asia is faster than in Europe; however Europe began extensive research on the benefits of migrating to IPv6 in January 2014. The USA use the new version with the Internet mobility, and the US Department of Defense (DoD) led other companies that that have a contract with them (such as IMB and Apple) to also switch to IPv6 [16]. On the other hand, there are over 200 million users in China that should be supported by IPv6; in 2008 the Olympic event began and this was the catalyst to push towards a move to IPv6, in order to provide modern networks which are able to cover everything from CCTV to taxis. Hong Kong Shanghai Banking Corporation's (HSBC) migrated to IPv6 to keep pace with development. The Australian government built a migration strategy in 2007 to move to IPv6 in 2012: it required an upgrade to all software and hardware to IPv6. Europe will provide the new version to all customers at the end of 2015: many of its economic organizations and enterprises prepared a report to demonstrate the importance of migration. In North America, the Obama administration on September 30, 2012 said that all web services, domain name systems (DNS), email, and other applications should move to IPv6 before the end of September 30, 2014. According to New Zealand's Chief Information Officer (CIO survey, the cost is still the primary factor for many organization in avoiding the migration; however, the same survey shows the IPv6 use is growing from 54% in 2009 to 74% in 2010. Microsoft has begun to provide applications that depend on IPv6, such as Remote Assistance in Windows 7 and Direct Access in both Windows 7 and Windows Server 2008 [17].

## 5. OPNET MODELER

The Optimized Network Engineering Tools (OPNET) Modeler is an efficient way to provide a complete study for the network analysis. The graphical user interface (GUI) is simple to use and the result is shown as graphical and static. Furthermore, it does not require a programing knowledge, and this can be easily used. The Opnet analyses the network as a real life network which gives a complete view before building the network in a real life and also contains a library of protocols and models which can be used as examples [18].

There are four simulation technologies supported by Opnet:

**Discrete-event Simulation (DES):** The DES provides a simulation in the same way as a real network; it will assist the study by analyzing the performance and behavior of the protocols and packets.

**Hybrid Simulation:** Hybrid simulations supported by DES, the results depend on analysis and the DES to be accurate; by using the two, accurate results can be generated with reasonable runtimes.

**Flow Analysis**: Analytical techniques and algorithm are used with flow analysis. It uses detailed configuration information to recreate the routing table for the device to ensure high efficiency. Flow analysis is used to study and understand the routing and reachability throughout the network.

**ACE Quickpredict**: The bandwidths, packet loss and latency impact time will be studied by ACE quick predict, which is supported within the OPNET Application Characterization Environment (ACE) [19]. Figure 11 shows the options for the Opnet simulation

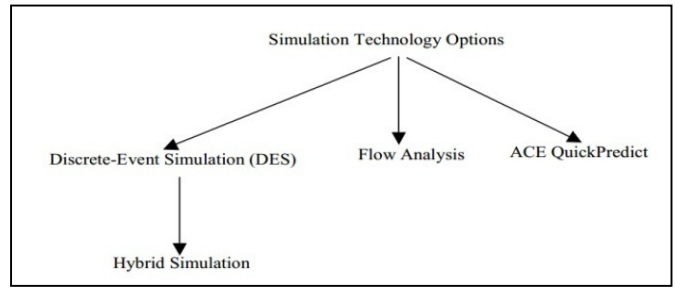

Figure 11. Opnet Simulation technology options

## 6. SIMULATION EXPERIMENT

This experiment will be conducted using the Opnet Modeler 17.5 for simulation. Opnet was chosen as it is considered as highly efficient simulation software and will be appropriate in reaching the goal of the experiment. It also includes most of the network technology such as routers and switches, as well as other equipment such as the filters, which help to analyses the traffic. The experiment consists of different stages: firstly, the network model is created; secondly, the most suitable statistical analysis is detected; thirdly, the simulation is run in order to obtain results; fourthly, the results are analyzed and compared to each other, and finally a report is written to discuss the results. .

The experiment will show the performance of two types of IPv6 transitions (6to4 and manual) and both of them will be compared with IPv4 and IPv6 traffic. The two transition technologies will provide the connection for two IPv6 sites that are isolated by an IPv4 network. The design includes a LAN, which contains 100 users that are connected by four switching devices using 100 base T cable. The Dual-Stack router (the router in the middle of the two networks) has two interfaces, one connected to the LAN by 100 Base T cable and the other to an IPv4 network by a PPPDS3 cable. The IPv4 network contains five routers connected with each other by 100 Base T links. The other side consists of the IPv6 network including a web server; the web server connection can be used for heavy browsing. This experiment will show how 6to4 and manual tunnel can be used to provide a connection for an IPv6 isolated by an IPv4. The routing protocol is RIP for IPv4 and RIPng for IPv6. The OSPFv3 is not compatible with the 6to4 tunnel. In addition, the experiment network topology is designed to be easy to understand and explains how the transition works. Moreover, the results will show the network behavior, which can be analyzed and compared with IPv4 and IPv6.

## 7. METHODS USED

The network is implemented by using different network materials such as switches, routers, LAN and servers as shown in figure 12. From the object palette one can choose the equipment and connect them together.

After the network implementation, start to configure the attributes for IPv6, IPv4, Dual-Stack, Manual and 6to4. The IPv4 and IPv6 they do not require any special configuration; just a normal configuration by assigning all interfaces IPv4 address for IPv4 network and IPv6 addresses for IPv6 network. The Dual-Stack as well is just required for each interface to add both IPs version 4 and version 6 together that is for servers and routers. The LAN setting can be configured by right clicking on the LAN icon and choosing "edit attributes." Extend "IP" and then "IP Host Parameters." IPv4 is assigned by putting its IP address in "Address" and then putting the mask in "subnet" mask. These steps should be followed again for each LAN.

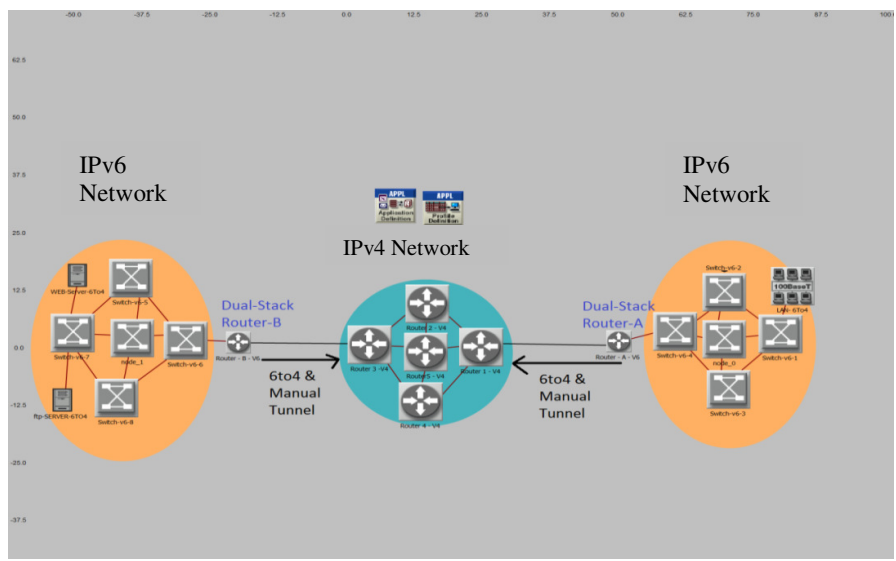

Figure 12.  The network topology in OPNET

The router setting is edited by clicking on "IP Routing Parameter" and from there "Interface Information." There, IP addresses for 12 ports for IPv4 can be assigned, and IPv6 addresses may also be added by clicking on "IPv6 Parameter". The numbers of ports depend on the number of addresses. Change the "not active" to "Default EUI-64" and assign the IPv6 to the network. The switch does not require any configuration, and it is important to choose the right cable with the right equipment. All nodes are connected together by 100 base T cables and the edge routers connect to the IPv4 backbone by PPP_DS3 cable.

There are two important factors that should be configured to establish the traffic when designing the network. First "Application" is used to decide which application is required for the network, by choosing it from the "Object Palette" and right clicking to extend the "Application Definitions." From the "Number of Rows" choose the number of applications that will apply for the network; in this experiment it is one (web server). The service types are assigned in "Application Definitions," and this information is relayed back to the server. The server can then decide which services will be applied to the network. The "Description" field gives the choice as to what type of service and the value; for this experiment the choice is HTTP and the value is heavy browsing, which includes videos and images. "Profile" is used to detect the properties for each application: "Application" is binding with the server and "Profile" is binding with the host. The Opnet Modeler provides many statistics which can be used to measure the network, which assists in understanding network behavior. The table 1 give examples of these measurements include:

Table 1. Opnet Modeler Parameters

| Download Response Time | The time required to send and receive a request from the client to the server. |
|---|---|
| Upload Response Time: | The required time to get an acknowledgement from the application. |
| Throughput | Represents the successful transmission of packets between two nodes. |
| Traffic Receiver (byte/sec): | Calculates the average bytes per second sent from client to server |
| Traffic received (packet/sec): | Calculates the number of packets by getting the average throughput of the packets sent from the client to the server. |
| Traffic Sent (byte/sec): | Is the average number of bytes assigned to the client by the server. |
| Traffic Sent (packet/sec): | Is the average number of packets assigned by the server to the client per second. |
| Ethernet Delay | Calculates the end-to-end delay between devices. |
| Utilization | Shows the percentage of consumption delay for packet forwarding and processing. |

**7.1 6to4 Tunneling**

To create a 6to4 tunnel, the IPv4 address is embedded between the middle router and the IPv4 with the IPv6 6to4 prefix address (2002:ipv4 address- with the edge router::/48). The middle router should be a dual stack router to support both versions of IPs and is considered the transport gate to the other site.

Table 2. 6to4 configurations

| LAN 6to4 | IP Address: 2002:192.168.1.1:1::2 |
|---|---|
| Dual-Stack Router- A | IPv4 192.168.1.1<br>IPv6 2002:192.168.1.1:1::1 |
| Tunnel configuration | Tunnel type: 6to4<br>Tunnel Source: IF10<br>Address 2002:192.168.1.1:d::1<br>Prefix: 128 |
| Dual-Stack Router- B | IPv4 10.1.1.1<br>IPv6 2002:10.1.1.1:a::1 |
| Tunnel configuration | Tunnel type: 6to4<br>Tunnel Source: IF10<br>Address 2002:10.1.1.1:b::1<br>Prefix: 128 |

The tunnel configuration is done from the "IP Routing Parameters." Choose the "Tunnel Interface," and from "Tunnel Information" detect the source, which is the interface that connects to the IPv4 network (Tunnel Source). From "Tunnel Mode" detect the tunnel type: for this experiment the tunnel type is 6to4. To assign the tunnel address choose "Tunnel Interface" from "IPv6 Parameters", extend the row for however many tunnels there are, and then assign the IP address that is the IP address for the tunnel. For the other site it is the same configuration but with different IPs. Table 2 shows the configurations for 6to4 tunnel; The tunnel is now ready to transfer packets through the IPv4 network.

### 7.2 IPv6 Manual Tunnel

The manual tunnel is not too different from the 6to4; the idea is similar, but for the tunnel mode choose IPv6 (Manual). The source is the interface with central edge router that connects to the IPv4 network, and the destination is the interface for the IPv6 on the other side (across the IPv4 network). The configuration is similar on the other side but notice the difference for IPs: manual strategy uses 2001 instead of 2002 for 6to4. Also, manual requires the source and destination for the tunnel as it works point-to-point, unlike 6to4 which works as point-to-multipoint. Table 3 shows the configurations for the manual tunnel.

Table 3. Manual configurations

| LAN manual | Address  2001:192.168.1.1:1::2 |
|---|---|
| Dual-Stack Router- A | IPv4 192.168.1.1<br>IPv6 2001:192.168.1.1:1::1 |
| Tunnel configuration | Tunnel Type: manual<br>Tunnel Source: IF10<br>Tunnel Destination: 10.1.1.1<br>Address 2002:192.168.1.1:d::1<br>Prefix: 128 |

| Dual-Stack Router- B | IPv4 10.1.1.1<br>IPv6 2001:10.1.1.1:a::1 |
|---|---|
| Tunnel configuration | Tunnel Type: manual<br>Tunnel Source: IF10<br>Tunnel Destination: 192.168.1.1<br>Address 2002:10.1.1.1:b::1<br>Prefix: 128 |

## 8. RESULTS AND DISCUSSION

The simulation ran for 5min (300 sec): this time is enough to gain an overview of the behavior of the network.

**8.1 Delay**

Figure 13 shows the LAN TCP delay within the web server: 6to4 is represented by green, manual transition by red, blue for the IPv4, and purple for IPv6.

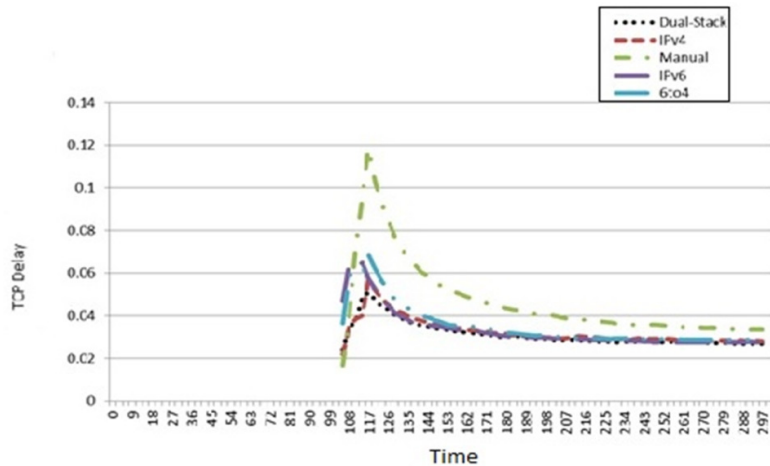

Figure 13. TCP network delay

The TCP delay occurred in the transport layer; the transport layer could use UDP or TCP protocol, as the connection with the web server requires a TCP connection. The transport layer is responsible for the connection between end to end and for the flow packets, so when delay occurs it is found in this layer. As seen from the graph, IPv6 and IPv4 have shorter delays than the transition strategies. Moreover, the delay occurs because of the congestion in the network. This network includes a LAN of 100 hosts; at the same time as the transmission was underway, each host was browsing, including accessing large images and long videos, which generated a high numbers of packets travelling inside the network and it is this which caused the congestion to occur. As previously discussed, the simulation programmer creates traffic by sending the same request from the server to the 100 hosts at the same time. This traffic causes congestion and therefore the delay is appears; after the requests are sent, the windowing is decreased, which therefore decreases the traffic flow and therefore minimizes delay.

In addition, the transition technologies caused further delay, because when the packets reached the middle router (between the IPv4 network and the IPv6 network), the packets transferred

33

through the tunnel. This required the IPv6 packet to be encapsulated into Ipv4 in order for them to transfer through the IPv4 infrastructure. When they reached the other side these packets were de-capsulated and transferred to the destination server; the time between capsulation and de-capsulation generated additional delays. Moreover, the transition technologies generated additional size to their packets (20 byte adds more than the original size). In addition, when the queue was full there was not enough space for the new packets; this generated drop packets and, as TCP is a reliable connection, when packets are dropped, TCP will retransmit the dropped packets. There are many factors that generate congestion, such as:

- **Queuing delay:** The packets arriving at the switch or the router will wait in the queue for processing and the waiting time will create a delay. Figure 14 shows the point to point delay between the dual stack router (middle router between two networks) and the IPv6 switch site. The queue delay increases with time because as the packets begin to reach the queue router and await their turn for processing, more packets are arriving: when the queue is full this will affect the network's throughput as the number of packets that are successfully arriving at their destination is decreasing. The figure shows IPv4 has less queue delay because the packet size is smaller than other types.

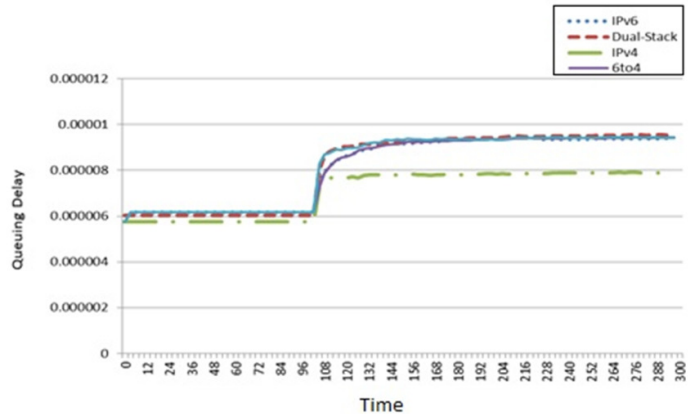

Figure 14. Queuing delay between point to point

As seen in the figure, the IPv4 and IPv6 have the lowest delay, because the processing time for the router is faster than for manual and 6to4 strategies. In addition, IPv4 has lees queue than IPv6 because the packet size for IPv4 less than IPv6.

- **CPU Utilizations**: The router contains a CPU, and the usage becomes extremely high when there is a high level of traffic that needs to be processed quickly. The manual and 6to4 transitions generate more pressure on the CPU than IPv4, IPv6 and Dual-Stack because the 6to4 and manual required the encapsulation and De capsulation operations on each packet cross from IPv6 to Ipv4 and vice versa; figure 15 shows the effect of the five phases on the router. In general, the CPU utilization represents the percent of CPU time spent in processing traffic (Chen, Chang, Lin, 2004).

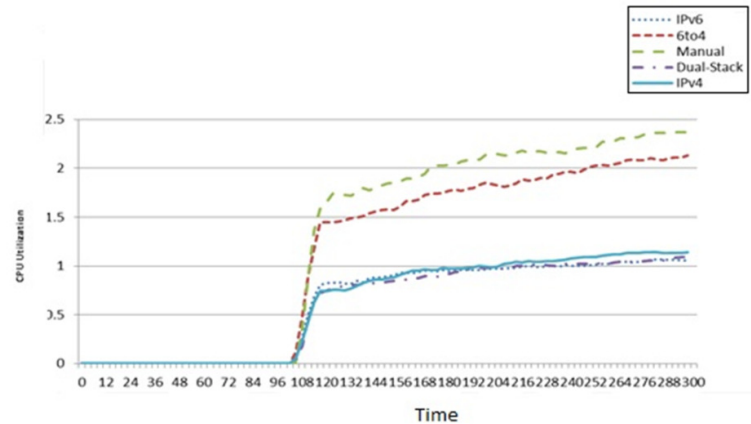

Figure 15. CPU utilization on the middle router

- **Page response time**: The response time for the page is related to the delay, as delay can influence the time it takes for the server to respond the host request. Figure 16 shows the responses times for each phase. The response times for IPv4, IPv6 and Dual-Stack are less than those of 6to4 and manual as they generated less overhead on the router and therefore the processing time is faster. The manual transition had a slower response time than 6to4, which is why the TCP delay is higher.

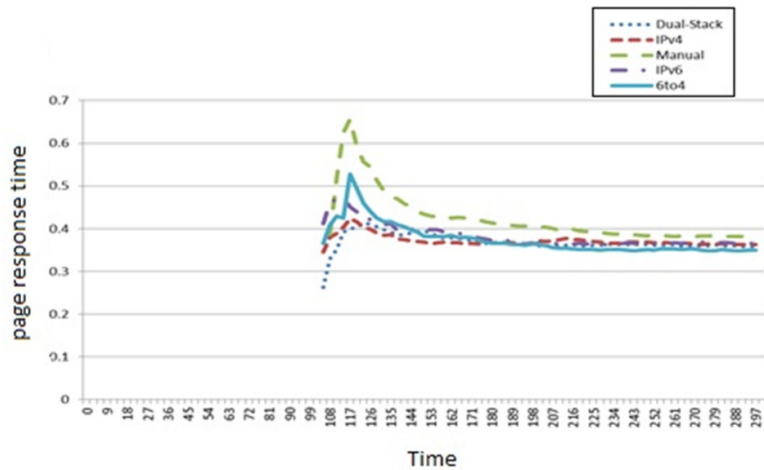

Figure 16. Server response time

Towards the end, congestion was generating the delay. For example, if there are 1000 packets that require 1 second to transmit in a normal situation with congestion the time needed can at least double, due to each packet needing to wait for a longer period of time, especially with TCP because of its reliability. The table 4 shows the difference in delays between IPv4, IPv6, manual and 6to4 tunneling. Table 4 shows the average network performances

Table 4. Average Network Performances

| Statistic<br>Phase | Page response time | TCP delay | Queue Delay | CPU Utilization |
|---|---|---|---|---|
| **Dual-Stack** | 0.3379 | 0.0375 | 7.7897 | 0.5453 |
| **IPv4** | 0.3852 | 0.0392 | 6.8323 | 0.5706 |
| **Manual Tunnel** | 0.5186 | 0.0677 | 7.6016 | 1.1845 |
| **IPv6** | 0.4165 | 0.0461 | 7.5914 | 0.5348 |
| **6to4 Tunnel** | 0.4381 | 0.0485 | 7.595 | 1.0651 |

### 8.2 Throughput

Throughput is the rate of transferring data through a network and is calculated using bits per second. The table shows the throughput between the LAN, which contains 100 users, and a FTP server. There are three data rates for this test: 1Mbps, 2 Mbps and 5 Mbps. The IPv6 had a higher throughput due to its larger packet size in comparison to the others; this enabled it to transfer more data. The second better throughput is for manual tunnel for 1 and 2 Mbps. With 5 Mbps the 6to4 got higher throughput than the others, except IPv6. Manual tunnel is point to point connection, which meant that the data travelled immediately from source to destination (which is also considered preferable for security reasons). In general, the throughput is increasing when the data rate is increased. Table 5 measure the throughput by using 1, 2 and 5 Mbps.

Table 5. Average Networks Throughput

| Phase | Date rate | Throughput (byte/sec) | Date rate | Throughput (byte/sec) | Date rate | Throughput (byte/sec) |
|---|---|---|---|---|---|---|
| **IPv6** | 1 Mbps | 19522.5 | 2 Mbps | 36707.5 | 5 Mbps | 83013 |
| **IPv4** | 1 Mbps | 13698.5 | 2 Mbps | 28553.5 | 5 Mbps | 74279.5 |
| **Dual-Stack** | 1 Mbps | 17775 | 2 Mbps | 32631.5 | 5 Mbps | 74275 |
| **Manual Tunnel** | 1 Mbps | 18356 | 2 Mbps | 33796 | 5 Mbps | 61167.5 |
| **6to4 Tunnel** | 1 Mbps | 15737.5 | 2 Mbps | 33213.5 | 5 Mbps | 75731 |

## 9. CONCLUSION

In the previous years IPv4 has proven its worth in providing sufficient addresses for the Internet. When the Internet continued to expand, it began to approach IPv4's limit in providing different services and applications. Therefore, a new version of IP (namely IPv6) was developed in order to cater to all users' requirements. In this paper, some networks were designed and simulated by using Opnet Modeler to study different translation schemes. The design contained different network devices in order to capture a real network environment. The network topology was configured in five phases as - IPv4, IPv6, Dual-Stack, 6to4, and manual tunnel. The statistical analysis was done to provide suitable results and to show that the network's performance varied

International Journal of Computer Networks & Communications (IJCNC) Vol.6, No.5, September 2014

across different mechanisms. For example, the CPU utilization for manual and 6to4 is higher than IPv6, IPv4 and Dual-Stack because the transition technology generates more effort to encapsulate and decapsulate. The Dual-Stack found less delay with TCP, but with 6to4 and manual the delay is higher because the packets are not transferred directly, as usual. The throughputs of the four network simulations were analyzed by using three different data rates: 1, 2 and 5 Mbps. The results show that IPv6 has higher throughput than the other four, and for manual it is higher than 6to4 till 5 Mbps. The 6to4 and manual strategies required manual configurations to detect the source, and the manual tunnel is required to have the destination detected in order to build the point to point mechanism.

**Authors**

Ali Albkerat received the B.E degree in Computer engineering from the 7[th] of April University, Libya in 2008 and has graduated MSc degree in Network Systems from the School of Computing, Teesside University, UK in 2014.

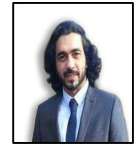

Dr Biju Issac is a senior lecturer in the School of Computing, Teesside University, UK. He has Bachelor of Engineering in Electronics and Communication Engineering (ECE), after which he completed a Master of Computer Applications (MCA) with honours. Later he finished his PhD in Networking and Mobile Communications, by research. He is a Charted Engineer (CEng), and Senior Member of IEEE.

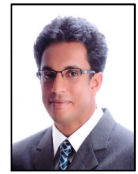